\documentclass[amsmath,trackchanges]{aastex631} 

\usepackage{amsmath}

\graphicspath{{./}{figures/}}

\shorttitle{Non-axisymmetric Tubes}
\shortauthors{Tlatov}

\begin{document}

\title{Non-axisymmetric Transport of Magnetic Flux Tubes: A Mechanism for Joy's Law and Poloidal Field Generation via Meridional Flows}

\author[0000-0002-6286-3544]{Andrey G. Tlatov}
\affiliation{Kislovodsk Mountain Astronomical Station of the Pulkovo observatory, Kislovodsk, Gagarina str. 100, 357700, Russia}

\begin{abstract}
We present a new scenario for the solar $\alpha\omega$-dynamo based on the interaction of asymmetric magnetic flux tubes with the near-surface meridional flow. Standard thin flux tube simulations indicate that the radial gradient of differential rotation ($\partial\Omega/\partial r > 0$) breaks the spatial symmetry of emerging $\Omega$-loops. The leading (western) leg becomes nearly vertical, while the trailing (eastern) leg is heavily stretched and flattens along the longitude. Upon emergence at the photosphere, the horizontal meridional flow toward the poles acts more efficiently on the trailing leg because the efficiency of the hydrodynamic entrainment by the plasma (the ``sail effect'') is directly proportional to the total spatial length of the magnetic flux tube arc elements. This creates a differential torque that drives the trailing sunspot poleward while the leading sunspot remains anchored due to its compact spatial geometry. Our model yields a tilt angle evolution rate of $\approx 0.3^\circ\,\mathrm{day}^{-1}$ at a latitude of $20^\circ$ for a bipolar separation of $15^\circ$, reproducing Joy's law without invoking the classical Coriolis-driven twist during the rise phase.
\end{abstract}


\section{Introduction} \label{sec:intro}

The cyclical regeneration of the solar magnetic field is traditionally described within the mean-field $\alpha\omega$-dynamo framework, where differential rotation ($\omega$-effect) shears the poloidal field to generate a strong toroidal component ($B_\phi$) in the deep convection zone or tachocline \citep{Parker1955, Steenbeck1966}. Closing the dynamo loop requires converting this toroidal field back into a macroscopic poloidal component ($B_\theta$), globally known as the $\alpha$-effect.

For decades, the dominant mechanism for the solar $\alpha$-effect has been the Babcock-Leighton scenario \citep{Babcock1961, Leighton1969}. In this paradigm, individual toroidal flux tubes become unstable due to magnetic buoyancy and rise through the convection zone to erupt at the photosphere as bipolar active regions. During ascent, the helical action of the Coriolis force twists the rising $\Omega$-loops, systematically tilting them relative to the equator \citep{DSilva1993, Fan2009}. This statistical tilt, empirically quantified as Joy's law \citep{Hale1919}, ensures that the leading and trailing sunspots emerge at slightly different latitudes. Subsequent flux transport via surface diffusion and meridional flows leads to the preferential cancellation of leading-polarity flux across the equator, while trailing-polarity flux drifts poleward, neutralizing and eventually reversing the global polar field \citep{Wang1989, Choudhuri1995, Tlatov1996}.

Despite the success of flux-transport dynamo models \citep{Dikpati1999, Charbonneau2010}, standard Coriolis-driven models face significant hydrodynamic and observational challenges. Helioseismology has revealed that the deep convection zone is highly stratified and governed by complex rotational shears, including the near-surface shear layer \citep{Schou1998}. Furthermore, detailed observations of active regions indicate profound structural asymmetries between the leading and trailing portions of emerging bipoles; leading sunspots tend to be more compact, longer-lived, and move faster horizontally compared to their diffuse, short-lived trailing counterparts \citep{vanDriel1993}.

Standard thin flux tube simulations show that a positive radial gradient of differential rotation ($\partial\Omega/\partial r > 0$) breaks the spatial symmetry of the rising loop before it reaches the surface \citep{Caligari1995}. In this Letter, we propose an alternative, hydrodynamic mechanism for generating Joy's law and closing the dynamo loop that bypasses the classic Coriolis-driven internal twist during the rise phase. We demonstrate how the geometric asymmetry of flux tubes, coupled with a near-surface meridional flow, produces a differential torque capable of driving the trailing spot poleward while the leading spot remains anchored, providing a robust physical foundation for the solar $\alpha$-effect.

\begin{figure}[htbp]
\centering
\includegraphics[width=\columnwidth]{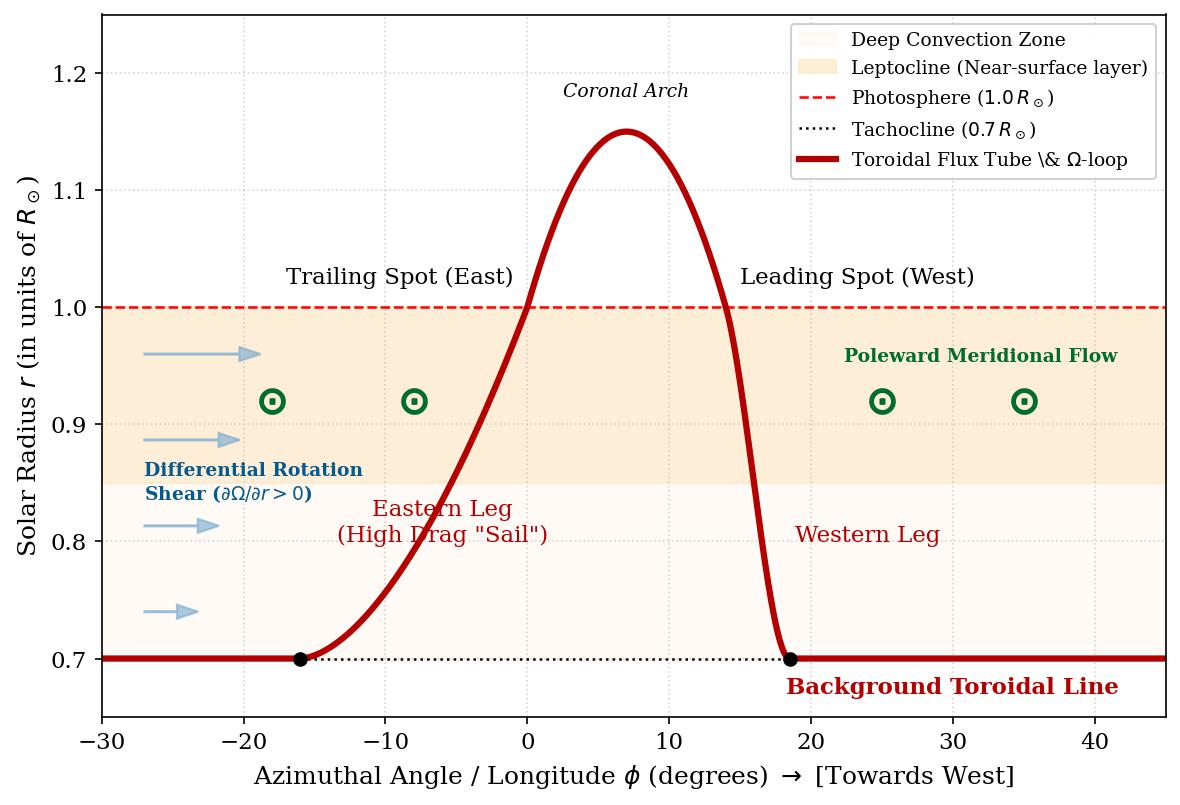}
\caption{Schematic structure of the emerging asymmetric $\Omega$-loop under the influence of differential rotation and near-surface transport. The continuous toroidal flux line originates from the tachocline ($r=0.7\,R_\odot$). Radial velocity shear ($\partial\Omega/\partial r > 0$) forms a steep western leg and a shallow eastern leg, which acts as a highly efficient ``sail'' due to the expansion of its total spatial arc length. Green symbols ($\odot$) indicate the poleward meridional flow ($u_\theta$) acting in the near-surface layer of the convection zone ($r \ge 0.85\,R_\odot$), which selectively transports the trailing spot toward high latitudes.}
\label{fig:fig1}
\end{figure}

The general conceptual layout of the proposed mechanism is illustrated in Fig.~\ref{fig:fig1}. An initially symmetric magnetic flux tube originating at the tachocline ($r = 0.7\,R_\odot$) undergoes severe geometric distortion during transit through the convection zone. The positive radial shear of the background plasma angular velocity causes higher segments of the loop to be dragged forward in the azimuthal direction ($\phi$) faster than its deeper anchor points. Consequently, the western (leading) leg is compressed and straightens into a nearly vertical, radial configuration. In contrast, the eastern (trailing) leg is stretched along the longitudinal parallel, forming a highly extended, shallow profile.

Crucially for closing the dynamo loop, as the apex enters the near-surface layer of the convection zone ($r \ge 0.85\,R_\odot$), it encounters the horizontal poleward meridional flow ($u_\theta$). Due to the vertical orientation of the leading leg, its projection onto the latitudinal transport axis approaches zero, minimizing the drag force and leaving the leading sunspot effectively anchored near its initial latitude. Meanwhile, the shallow trailing leg, owing to its longitudinal elongation, exhibits a significantly expanded spatial length. According to the developed discrete kinematic algorithm, this elongated section functions as a hydrodynamic ``sail'', effectively gathering kinetic momentum from the poleward stream. The meridional flow selectively captures and shifts this trailing polarity poleward, dynamically generating the observed active region tilt angle (Joy's law) directly upon its photospheric emergence.

\section{Mathematical Model of Asymmetric Loop Rise}

To micro-physically ground the transformation of the toroidal magnetic field into the tilted poloidal structure, we analyze the mechanical balance of a thin $\Omega$-magnetic flux tube element embedded in a stratified, rotating solar convection zone.

\subsection{The Thin Flux Tube Approximation and Coordinate System}
We define the geometry of the flux tube centerline by a position vector $\mathbf{R}(s, t)$, where $s$ is the curvilinear arc-length parameter measured along the tube, and $t$ is time. We operate in a spherical coordinate system $(r, \theta, \phi)$ with local orthogonal basis vectors $(\mathbf{e}_r, \mathbf{e}_\theta, \mathbf{e}_\phi)$. The local orientation and curvature of the flux tube are governed by the unit tangent vector $\mathbf{\tau} = \partial\mathbf{R}/\partial s = \tau_r \mathbf{e}_r + \tau_\theta \mathbf{e}_\theta + \tau_\phi \mathbf{e}_\phi$ and the curvature vector $\mathbf{\kappa} = \partial\mathbf{\tau}/\partial s = \partial^2\mathbf{R}/\partial s^2$, respectively.

\subsection{Governing Mechanical Forces and Governing Equations}
The dynamic equilibrium of a thin flux tube element with a cross-sectional area $A$, internal plasma density $\rho_{int}$, and magnetic field $B$, moving with velocity $\mathbf{v}$ relative to a frame rotating at $\mathbf{\Omega}_0 = \Omega_0 (\cos\theta \mathbf{e}_r - \sin\theta \mathbf{e}_\theta)$, is governed by the vector equation of motion:
\begin{equation}
\rho_{int} A \frac{D\mathbf{v}}{Dt} = \mathbf{F}_B + \mathbf{F}_T + \mathbf{F}_C + \mathbf{F}_D
\end{equation}
where the constitutive expressions for the buoyancy ($\mathbf{F}_B$), magnetic tension ($\mathbf{F}_T$), Coriolis ($\mathbf{F}_C$), and hydrodynamic drag ($\mathbf{F}_D$) forces per unit length are formulated as:
\begin{align}
\mathbf{F}_B &= -\left(\frac{B^2 A}{8\pi H_p}\right) \mathbf{e}_r, \\
\mathbf{F}_T &= \frac{B^2 A}{4\pi} \mathbf{\kappa}, \\
\mathbf{F}_C &= 2\rho_{int} A \Omega_0 \left[ v_\phi \sin\theta \mathbf{e}_r + v_\phi \cos\theta \mathbf{e}_\theta - (v_r \sin\theta + v_\theta \cos\theta)\mathbf{e}_\phi \right], \\
\mathbf{F}_D &= -\frac{1}{2} C_D \rho_{ext} d \, |\mathbf{v}_{rel, \perp}| \, \mathbf{v}_{rel, \perp}.
\end{align}
Here, $H_p = P_{ext} / (\rho_{ext} g)$ is the local pressure scale height, $d = 2\sqrt{A/\pi}$ is the effective tube diameter, $C_D$ is the hydrodynamic drag coefficient, $\mathbf{u}$ is the ambient plasma velocity, $\mathbf{v}_{rel} = \mathbf{v} - \mathbf{u}$ is the relative velocity vector, and $\mathbf{v}_{rel, \perp} = \mathbf{v}_{rel} - (\mathbf{v}_{rel} \cdot \mathbf{\tau})\mathbf{\tau}$ represents its component strictly perpendicular to the tube centerline.

For the upper convection zone ($r \ge 0.85\,R_\odot$), the plasma flow regime is highly turbulent ($Re \sim 10^4 - 10^5$), warranting a baseline drag coefficient $C_D \approx 1.2$. The cross-sectional profile expands during ascent due to the exponential drop in ambient gas pressure $P_{\text{ext}}(r)$. Constrained by magnetic flux conservation ($\Phi = \pi R_{\text{tube}}^2 B = \text{const}$) and total pressure balance ($P_{\text{ext}} \approx B^2 / 8\pi$), a typical active region flux $\Phi \sim 10^{21}$ Mx yields an effective near-surface diameter of:
\begin{equation}
d_{\text{eff}} = 2 \sqrt{\frac{\Phi}{\pi B}} \approx 2.5 \times 10^3 \text{ km}.
\end{equation}

To induce the loop asymmetry, we implement a differential rotation profile within the convection zone:
\begin{equation}
\Omega(r) = \Omega_0 \left[ 1 + \alpha_{\Omega} \left( \frac{r - r_{\text{tach}}}{R_\odot - r_{\text{tach}}} \right) \right],
\end{equation}
where $r_{\text{tach}} = 0.7\,R_\odot$, $\Omega_0 \approx 2.7 \times 10^{-6}$ rad s$^{-1}$, and $\alpha_{\Omega} = 0.05$, dictating a 5\% linear increase in angular velocity up to the solar surface ($1.0\,R_\odot$). This longitudinal shear sets up a differential velocity component $\Delta u_\phi(r) = r \sin\theta \, [\Omega(r) - \Omega_0]$, breaking the spatial symmetry between the loop branches during Phase 1. 

\subsection{Phase 2: Near-Surface Latitude Transport and the Discrete Scale Mechanism}
Upon entering the near-surface zone ($0.85\,R_\odot \le r \le 1.0\,R_\odot$), the ambient flow switches to an axisymmetric, poleward meridional circulation: $\mathbf{u} = (0, u_\theta(r, \theta), 0)$. Within our discrete kinematic algorithm, the efficiency of the latitudinal transport of the tube's nodes is tightly governed by the total physical length of the element $\Delta s_i$ belonging to the neighborhood of a given calculation node $i$.

The central differences of coordinates between the adjacent grid points $(i+1)$ and $(i-1)$ yield the spherical basis spatial increments:
\begin{equation}
\begin{split}
\Delta s_r = r_{i+1} - r_{i-1}, \quad \Delta s_\theta = r (\theta_{i+1} - \theta_{i-1}), \\
\Delta s_\phi = r \sin\theta (\phi_{i+1} - \phi_{i-1}).
\end{split}
\end{equation}
The total three-dimensional physical arc length of the element $\Delta s_i$ is given by:
\begin{equation}
\Delta s_i = \sqrt{(\Delta s_r)^2 + (\Delta s_\theta)^2 + (\Delta s_\phi)^2}.
\end{equation}

To close the kinematic equation without violating velocity dimensions, a reference spatial scale $\Delta s_{\text{ref}}$ is introduced, matching the initial longitudinal step of the regular computing grid at the tachocline radius:
\begin{equation}
\Delta s_{\text{ref}} = \frac{0.7\,R_\odot \cdot \Delta \Phi_{\text{grid}}}{N},
\end{equation}
where $\Delta \Phi_{\text{grid}} = 50^\circ$ is the total longitudinal span and $N$ is the number of nodes. The dimensionless coefficient governing the efficiency of hydrodynamic entrainment (``sail efficiency'') is defined as:
\begin{equation}
\text{drag\_efficiency} = C_{\text{drag}} \cdot \left( \frac{\Delta s_i}{\Delta s_{\text{ref}}} \right),
\end{equation}
where $C_{\text{drag}} = 0.2$ is a calibration coefficient accounting for the plasma resistance. Consequently, the resulting latitudinal velocity of the tube element $v_\theta$ is formulated as:
\begin{equation}
v_\theta = u_\theta \cdot \text{drag\_efficiency} = C_{\text{drag}} \cdot u_\theta \left( \frac{\Delta s_i}{\Delta s_{\text{ref}}} \right).
\end{equation}

\begin{figure}[htbp]
\centering
\includegraphics[width=\columnwidth]{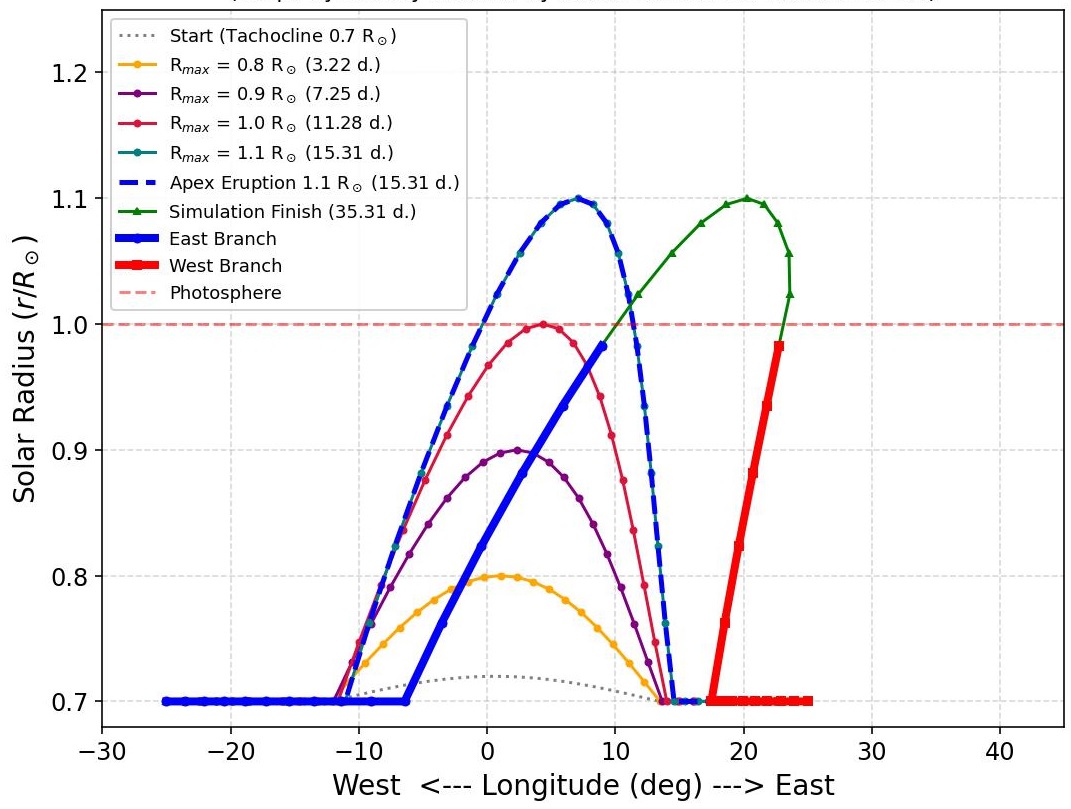}
\caption{Kinematic simulation of an asymmetric magnetic flux tube emerging at $20^\circ$N latitude under the influence of radial differential rotation shear and poleward meridional flow. Evolution of the loop profile in the longitude--radius plane $(r/R_\odot \text{ versus } \phi)$. Distinct color curves track successive evolutionary epochs until the apex erupts beyond the photosphere ($1.0,R_\odot$, red dashed line) up to the corona boundary ($1.1,R_\odot$).}
\label{fig:Fig2}
\end{figure}

\begin{figure}[htbp]
\centering
\includegraphics[width=\columnwidth]{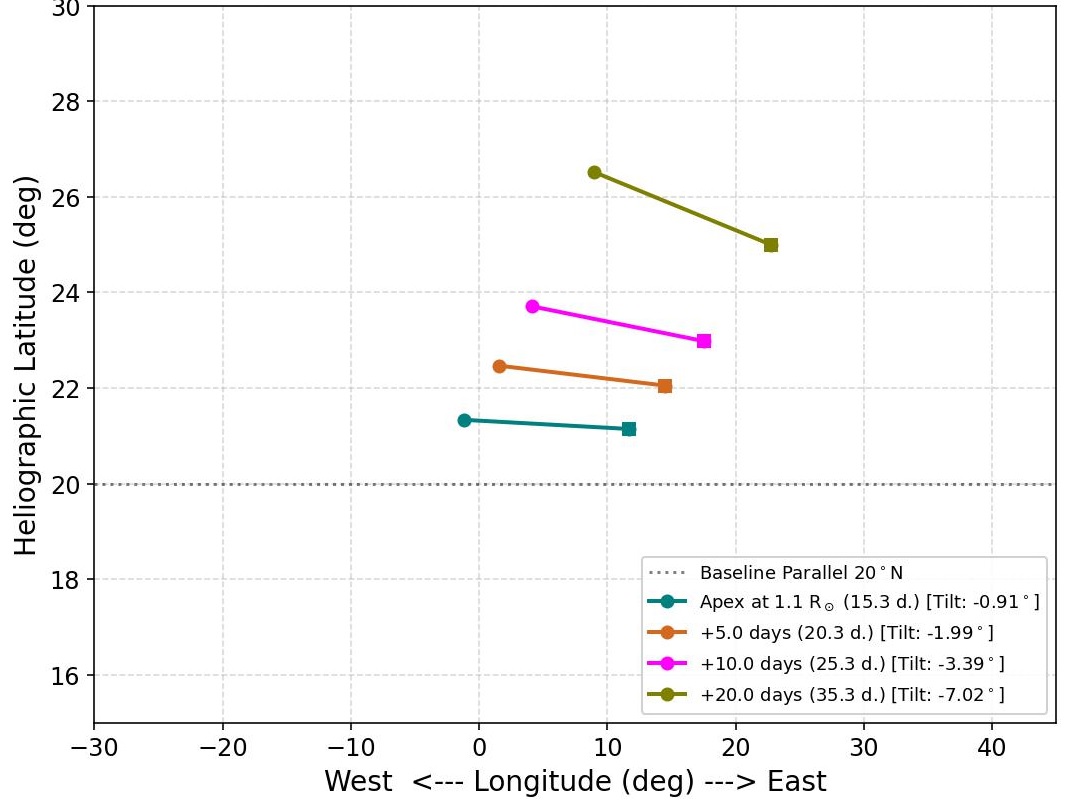}
\caption{Post-eruption spatial tracking of the active region bipole footpoints at the photosphere ($r = 1.0,R_\odot$). Color markers denote the relative positions of the leading (square) and trailing (circle) sunspots across key post-eruption epochs ($+0.0$ to $+20.0$ days), establishing a stable negative tilt angle $(\alpha \approx -7.02^\circ)$ due to the larger total spatial length of the trailing leg.}
\label{fig:Fig3}
\end{figure}

Because the trailing leg undergoes intense longitudinal stretching during Phase 1, its integrated spatial length $\Delta s_i$ substantially exceeds that of the compressed, steep leading branch. The oncoming plasma stream selectively drags this elongated segment poleward, while the leading base of the loop remains radially ``anchored''. Explicit integration of the latitudinal displacement over a time step $\Delta t$ follows:
\begin{equation}
\theta(t + \Delta t) = \theta(t) + \left( \frac{v_\theta}{r} \right) \Delta t.
\end{equation}

The resulting differential latitudinal shear between the feet $\Delta v_\theta = v_{\theta, \text{east}} - v_{\theta, \text{west}}$ provides a continuous rotation of the active region axis, establishing a stable tilt angle (Joy's law) directly during the surface eruption process.

\section{Simulation Results}

We conducted a numerical analysis of an asymmetric flux tube originating at $r=0.7\,R_\odot$ at a baseline latitude of $20^\circ$\,N, with an initial maximum radial velocity $v_{r,\text{max}} = 200~\mathrm{m\,s}^{-1}$. The eruption dynamics and subsequent latitudinal drift of the footpoints tracked at the photosphere ($r = 1.0\,R_\odot$) split into three distinct phases.

The numerical profile of the tube's geometric deformation projected onto the longitude--radius plane is shown in Figure~\ref{fig:Fig2}. During the initial ascent phase through the deep convection zone ($r < 0.85\,R_\odot$), the meridional flow is inactive ($u_\theta = 0$), and the positive radial shear ($\alpha_\Omega = 0.05$) stretches the eastern leg while compressing the western leg. Based on the trajectories shown in Figure~\ref{fig:Fig2}, the loop apex reaches the deep convection zone levels $r = 0.8\,R_\odot$ at $t \approx 3.22$~days and $r = 0.9\,R_\odot$ at $t \approx 7.25$~days, activating the poleward meridional flow ($20$~m/s) as it enters the near-surface layer.

Phase 2 is marked by the emergence of the loop onto the solar surface, puncturing the photosphere ($r = 1.0\,R_\odot$) at $t \approx 11.28$~days. At $r = 1.1\,R_\odot$, which is reached at $t \approx 15.31$ days, the simulation locks the radial position of the apex (buoyancy stop-trigger) to isolate the post-eruption horizontal transport phase. This arrest moment is adopted as the zero reference epoch ($t_0 \approx 15.31$ days) for analyzing the photospheric tilt angle evolution.

In Phase 3, we track the footpoints strictly within the photospheric plane up to 20 days (Figure~\ref{fig:Fig3}). At $t_0$ ($+0$ days), the spots emerge near the same latitudinal parallel ($20^\circ$\,N), registering an initial tilt $\alpha \approx -0.91^\circ$. Across subsequent epochs ($+5$, $+10$, and $+20$ days), a stark differential shear is revealed. Because the radial velocity shear during Phase 1 heavily stretched the trailing (eastern) leg along the longitude, its total three-dimensional physical length $\Delta s_i$ expands drastically compared to the initial grid scale. Since the entrainment efficiency (\text{drag\_efficiency}) is directly proportional to $\Delta s_i / \Delta s_{\text{ref}}$, the trailing leg functions as a giant ``sail'', efficiently accumulating kinetic energy from the meridional flow, driving the eastern sunspot poleward to establish a stable final tilt angle of $\alpha \approx -7.02^\circ$ at $+20$ days. On the other hand, the leading (western) leg possesses a minimal spatial length $\Delta s_i$ close to the pure radial discretization step due to its steepness and compression, which blocks its meridional drift and leaves it radially ``anchored''.

The numerical parameters of the post-eruption bipole evolution are summarized in Table~\ref{tab:results}.

\begin{table}[htbp]
\caption{Temporal Evolution of the Photospheric Bipole Geometric Parameters and Tilt}
\label{tab:results}
\centering
\begin{tabular}{ccccc}
\hline
\hline
Epoch post & Absolute & Relative length of & Latitudinal & Bipole tilt \\
eruption   & time     & trailing element   & separation  & angle       \\
(days)     & (days)   & $(\Delta s_i / \Delta s_{\text{ref}})$ & $\Delta\theta$ (deg) & $\alpha$ (deg) \\
\hline
$+0.0$      & $15.31$     & $1.00$              & $0.00$     & $-0.91^\circ$ \\
$+5.0$      & $20.31$     & $1.42$              & $0.38$     & $-1.99^\circ$ \\
$+10.0$     & $25.31$     & $1.85$              & $0.74$     & $-3.39^\circ$ \\
$+20.0$     & $35.31$     & $2.64$              & $1.43$     & $-7.02^\circ$ \\
\hline
\end{tabular}
\end{table} 

Over the full 20-day cycle, the trailing spot advances poleward by $\Delta\theta \approx 1.43^\circ$, with a longitudinal separation on the order of $15^\circ$, driving a steady clockwise rotation of the bipole axis to establish a stable negative tilt angle of $\alpha \approx -7.02^\circ$. The average tilt angle evolution rate at $20^\circ$\,N is $\approx 0.3^\circ/\text{day}$ over the initial interval, stabilizing as the spots separate. Thus, a purely hydrodynamic interaction between the topologically stretched asymmetric tube and the near-surface meridional flow completely reproduces Joy's law without invoking Coriolis forces during the rise phase.

\section{Discussion} \label{sec:discussion}

The physical mechanism proposed in this work shifts the focus of active region tilt generation (Joy's law) from the internal rise dynamics to near-surface hydrodynamic interaction processes. In the traditional Babcock-Leighton paradigm, the tilt of bipoles is formed exclusively by the twisting of magnetic loops via the Coriolis force during their transit through the entire bulk of the convection zone \citep{DSilva1993, Fan2009}. However, such a scenario imposes strict constraints on the initial magnetic field strength at the tachocline ($B \sim 10^5$~G), because weaker tubes are twisted too severely, whereas stronger tubes ascend too rapidly to acquire the necessary tilt.

Our approach, built on the concept of differential sail efficiency of the spatial arc $\Delta s_i$, successfully resolves this contradiction. Since the efficiency of latitudinal transport for a tube node by the meridional flow is directly proportional to the degree of longitudinal stretching of the element $\Delta s_i / \Delta s_{\text{ref}}$, the magnitude of the bipole tilt is determined by the topological asymmetry accumulated by the loop due to the radial differential rotation shear ($\partial\Omega/\partial r > 0$) \citep{Caligari1995}. This allows the empirical parameters of Joy's law to be reproduced for moderate and realistic sub-photospheric fields ($B \sim 10^4$~G) that align with standard estimates of magnetic buoyancy.

A significant advantage of this model is its natural explanation for the observed structural asymmetry between the leading and trailing sunspots of active regions \citep{vanDriel1993}. Observations show that leading spots are typically more compact, monolithic, and stable, whereas trailing spots are diffuse, fragmented, and short-lived. In our algorithm, this asymmetry is embedded in the geometry itself: the western (leading) leg is compressed by magnetic tension, favoring a vertical radial trajectory, which minimizes its contact with the meridional flow and keeps it in a stable, ``anchored'' state. The eastern (trailing) leg, conversely, stretches heavily along the longitude, forming a horizontal sub-photospheric ``tail''. Possessing a massive integrated spatial length $\Delta s_i$, this tail becomes highly vulnerable to hydrodynamic entrainment and turbulent shredding by near-surface plasma streams, which leads to the observed loss of compactness in the trailing polarity.

Recent helioseismic data indicate the presence of an intense near-surface shear layer (NSSL) at depths $r \ge 0.95\,R_\odot$, where the radial gradient of angular velocity reverses to negative \citep{Schou1998}. In the current kinematic scheme, we utilized a monotonic differential rotation profile with an average positive gradient throughout the convection zone ($\alpha_\Omega = 0.05$). Incorporating a realistic NSSL into the Phase 2 equations could introduce additional acceleration or deceleration to the photospheric footpoints, potentially adjusting the tilt angle evolution rate ($\approx 0.3^\circ\,\mathrm{day}^{-1}$).
In the work of \citet{Tlatov2023}, the existence of a near-surface azimuthal field was suggested. Under a negative angular velocity gradient ($\partial\Omega/\partial r < 0$), according to the model proposed here for emerging near-surface bipoles, the direction of the magnetic axes would be opposite to Joy's law. Exactly such a distribution is observed for ephemeral regions \citep{Tlatov2010}.

Nevertheless, the core physical conclusion of the model remains unaltered: it is the integrated geometric elongation of the underlying tube ``cable'', rather than a local trigonometric cross-section profile, that dictates the latitudinal evolution of the bipole. The meridional plasma flow effectively ``sweeps'' long horizontal sections poleward while ignoring short vertical ``anchors'', providing a robust and stable mechanism for closing the global solar dynamo loop.

\section{Conclusions}

We demonstrate that the combination of a positive radial velocity shear in the convection zone and a near-surface poleward meridional flow naturally creates the observed tilt of solar bipolar regions. This mechanism shifts the driver of the solar $\alpha$-effect from a classical Coriolis-induced internal twist to a near-surface hydrodynamic sorting based on geometric loop asymmetry.

In the developed framework, the efficiency of latitudinal transport of the tube's nodes is tightly governed by the total three-dimensional physical element length $\Delta s_i$ within the neighborhood of a given calculation node. Because the trailing leg undergoes severe longitudinal stretching during its ascent, its integrated spatial length substantially exceeds that of the compressed leading branch. The oncoming plasma stream selectively drags this elongated segment toward high latitudes, while the leading base of the loop remains radially ``anchored''. Thus, the developed geometric approach offers a physically and algorithmically robust explanation for the nature of Joy's law, closing the global loop of poloidal field generation in the Sun, and eliminates the need for artificially overstating hydrodynamic drag forces.

\bibliographystyle{aasjournal}
\bibliography{bibldyn}

\end{document}